\documentclass [a4paper,fleqn,12 pt]{article}
\usepackage{graphicx}
\usepackage {subfigure,epsfig}

\usepackage {amsmath} \usepackage{amssymb} \usepackage{cite} \usepackage{amsthm}
\numberwithin{equation}{section} \numberwithin{table}{section} \mathindent=0pt
\theoremstyle{plain} 

\numberwithin{theorem}{section}

\begin{document}

\title{Power solution expansions of the analogue to the first Painleve equation}
\author{Aleksandr D. Bruno and Nikolai A. Kudryashov}
\date{Department of Applied Mathematics\\
Moscow  Engineering and Physics Institute\\
(State university)\\
31 Kashirskoe Shosse,  115409\\
Moscow, Russian Federation} \maketitle

\begin{abstract}
\end{abstract}
{\footnotesize
\begin{quotation} The fourth-order analog to the first Painlev\'{e} equation is studied.
All power expansions for solutions of this equation near points
$z=0$ and $z=\infty$ are found. The exponential additions to the
expansion of solution near $z=\infty$ are computed. The obtained
results confirm the hypothesis that the fourth-order analog of the
first Painlev\'{e} equation determines new transcendental functions.
By means of the methods of power geometry the basis of the plane
lattice is also calculated.
\end{quotation}
}
\bigskip

\section{\!\!\!\!\!\!.\,\, Introduction.}

In \cite{Kudryashov01} the hierarchy of the first Painlev\'{e}
equation was suggested. It can be described by the relation
\begin{equation}
\label{1.1} L^n[w]=z
\end{equation}
where $L^n$ is the Lenard's operator, which is determined by the
relation \cite{Kudryashov02}

\begin{equation}
\begin{gathered}
\label{1.2} \frac{d}{dz} L^{n+1} = L_{zzz}^n -4wL^n_z - 2w_z L^n,
\,\,\quad\,\, L^0[w] =- \frac12
\end{gathered}
\end{equation}

Assumed that $n=0$  in \eqref{1.2}, we have $L^1[w]=w$. In case of
$n=2$ using \eqref{1.1} we get the first Painlev\'{e} equation
\cite{Golubev}

\begin{equation}
\label{1.4} w_{zz}-3w^2=z
\end{equation}

If $n=3$ in \eqref{1.1}, we get the fourth-order equation
\cite{Kudryashov01, Kudryashov02}

\begin{equation}
\label{1.5} w_{zzzz} -10w\,w_{zz}-5w^2_z+10w^3=z
\end{equation}

Using $n=4$,  we obtain the sixth-order equation from \eqref{1.1}

\begin{equation}
\label{1.5a.} w_{zzzzzz}-14ww_{zzzz}-28\,w_z\,w_{zzz}
-21\,w_{zz}^2+70\,ww_z^{2}-35\,w^4=z
\end{equation}

Equation \eqref{1.5} is used in describing of the waves on water
\cite{Olver, Kudryashov03} and in the Henon-Heiles model, which
characterizes the behavior of star in the middle field of galaxy
\cite{Henon, Fordy, Hone}.

In papers \cite{Kudryashov04, Kudryashov05, Kudryashov06,
Kudryashov07, Kudryashov08, Kudryashov09, Kudryashov10,
Kudryashov11, Joshi01, Mugan01, Cosgrove01, Gordoa01, Pickering01,
Clarkson01, Kawai01} it was shown that equation \eqref{1.5} has
properties, that are typical for  the Painlev\'{e} equations
$P_1\div P_6$. Equation \eqref{1.5} belongs to the class of exactly
solvable equations, as it has Lax pair and a lot of other typical
properties of the exactly solvable equations. However it doesn't
have the first integrals in the polynomial form, that is one of the
features of  the Painlev\'{e} equations. Equation \eqref{1.5} seems
to determine new transcendental functions just as equations $P_1\div
P_6$, although the rigorous proof of the irreducibility of equation
\eqref{1.5} is now the open problem.

Thereupon the study of all the asymptotic forms and power expansions
of equation \eqref{1.5} is the important stage of
 the analysis of this equation, as this fact indirectly confirms the
irreducibility of equation \eqref{1.5}.

Let's find all the power expansions for the solution of equation
\eqref{1.5} in the form of

\begin{equation}
\label{1.5.a} w(z)= c_r\,z^{r}+ \sum_{s}\,c_s\,z^{s}
\end{equation}
at $z\rightarrow\,0$, then $\omega=-1$, $s>r$ and at
$z\rightarrow\,\infty$, then $\omega=1$, $s<r$.

For that we use the methods of power geometry \cite{Bruno01,
Bruno02} by analogy with \cite{Bruno03}.

\section{\!\!\!\!\!\!.\,\,The general properties of equation \eqref{1.5}.}
Let's consider the fourth-order equation \eqref{1.5}

\begin{equation}
\label{1.7}f(z,w)\stackrel{def}{=}w_{zzzz} - 10ww_{zz} - 5 w^2_z +
10 w^3  - z =0
\end{equation}

For monomials of equation \eqref{1.7} we have points
$M_1=(-4,1),\,\,\, M_2=(-2,2),\,\,\,M_3=(-2,2),\,\,\,
M_4=(0,3),\,\,\, M_5=(1,0)$.

The carrier of equation is defined by four points $Q_1=M_1$,
 $Q_2=M_4 $,  $Q_3=M_5$  and  $Q_4=M_2=M_3$.  Their convex hull
$\Gamma$ is the triangle (fig. 1).

\begin{figure}[h] % было [p]
 \centerline{\epsfig{file=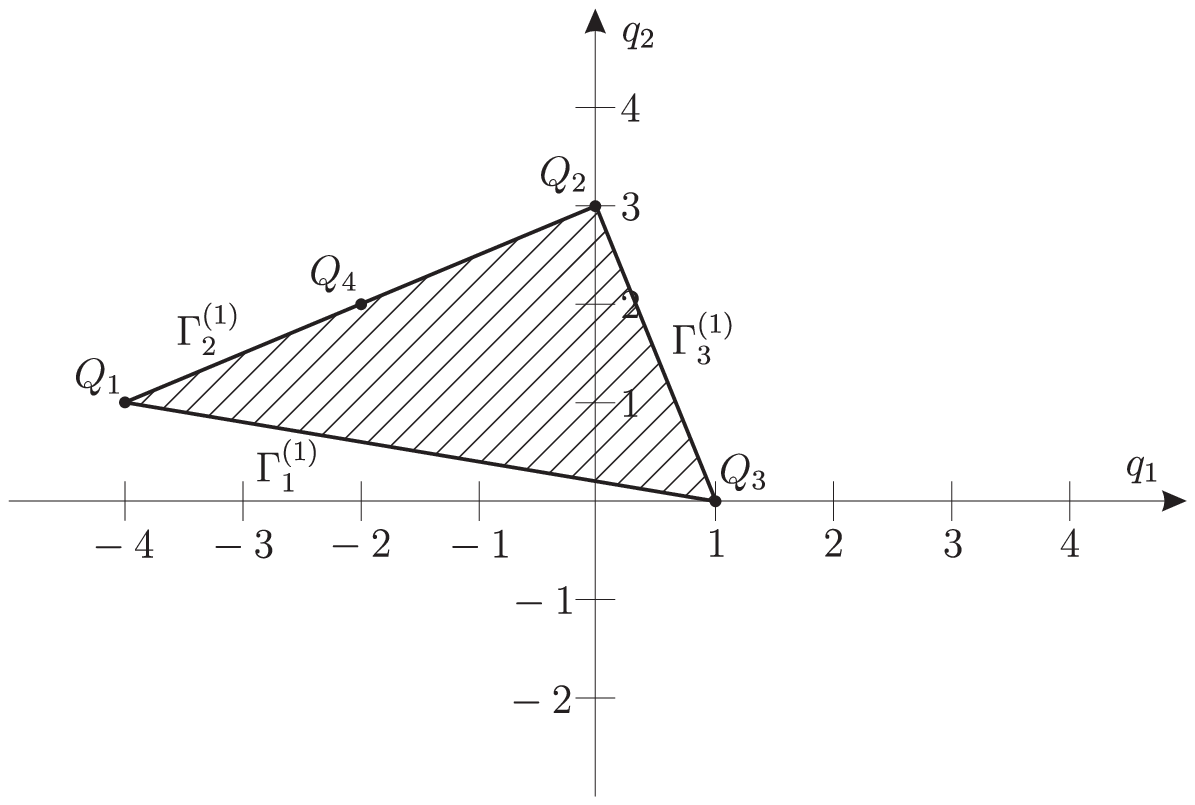,width=120mm}}
 \caption{}\label{fig:z_post_0}
\end{figure}

This triangle has apexes $Q_j\,(j=1,2,3)$ and edges
$\Gamma_1^{(1)}=[Q_3,\,Q_1],\,\, \Gamma_2^{(1)} =[Q_1,\, Q_2],\,\,
\Gamma_3^{(1)} =[Q_2,\, Q_3]$

Outward normal vectors $N_j\,(j=1,2,3)$ of edges $\Gamma_j^{(1)} \,
(j=1,2,3) $  are determined by vectors

\begin{equation}
\label{1.8}N_1=(-1,-5),\,\,\, N_2=(-1,2),\,\,\, N_3=(3,1)
\end{equation}

The normal cones $U_j^{(1)}$ to edges $\Gamma_j^{(1)}$ are

\begin{equation}
\label{1.9}U_j^{(1)} =\mu N_j,\,\,\, \mu>0,\,\,\, j=1,2,3
\end{equation}

They and the normal cones $U_j^{(0)}$ of apexes
$\Gamma_j^{(0)}=Q_j\,\, (j=1,\,2,\,3)$ are represented at fig. 2.

\begin{figure}[h] % было [p]
 \centerline{\epsfig{file=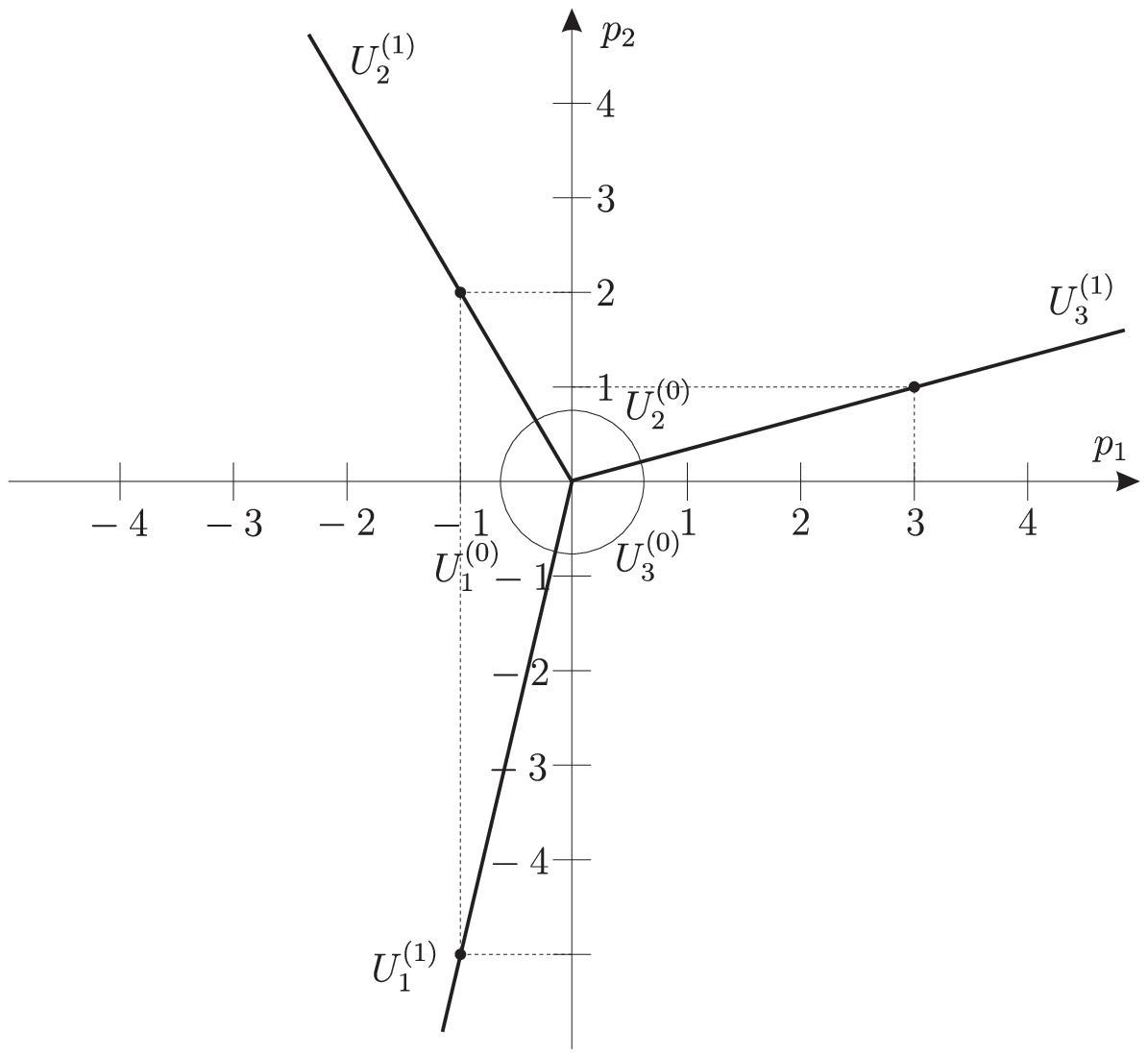,width=120mm}}
 \caption{}\label{fig:z_post}
\end{figure}

If the carrier of  equation \eqref{1.7} is moved by vector $-Q_3$,
then it is situated at the lattice $\textbf{Z}$, formed by vectors

\begin{equation}
\label{1.9a}Q_1 - Q_3=(-5,\,1),\,\,\quad\,Q_2-Q_1=(-1,\,3)
\end{equation}

We choose the basis of the lattice as

\begin{equation}
\label{1.9b}B_1=(-5,\,1),\,\,\quad\,B_2=(-3,\,2)
\end{equation}

Let's study solutions, corresponding to the bounds
$\Gamma_j^{(d)},\,\, d=0,1;\,\, j=1,\,2,\,3$ in view of the reduced
equations, conforming to apexes $\Gamma_j^{(0)} (j=1,\,2,\,3)$
\begin{equation}
\label{1.10}\hat{\emph{f}}_{1}^{(0)}\stackrel{def}{=}w_{zzzz}=0
\end{equation}
\begin{equation}
\label{1.11}\hat{\emph{f}}_{2}^{(0)}\stackrel{def}{=}10\,w^3=0
\end{equation}
\begin{equation}
\label{1.12}\hat{\emph{f}}_{3}^{(0)}\stackrel{def}{=}\,- z=0
\end{equation}
and reduced equations, conforming to edges $\Gamma_j^{(1)}
(j=1,2,3)$
\begin{equation}
\label{1.13}\hat{\emph{f}}_{1}^{(1)}\stackrel{def}{=}w_{zzzz}- z=0
\end{equation}
\begin{equation}
\label{1.14}\hat{\emph{f}}_{2}^{(1)}\stackrel{def}{=}w_{zzzz}
-10\,w\,w_{zz} -5w_z^2 + 10 w^3=0
\end{equation}
\begin{equation}
\label{1.15}\hat{\emph{f}}_{3}^{(1)}\stackrel{def}{=}10w^3- z=0
\end{equation}

Note, that the reduced equations \eqref{1.11} and \eqref{1.12} are
the algebraic ones. According to \cite{Bruno02} they don't have
non-trivial power or non-power solutions.

\section{\!\!\!\!\!\!.\,\, Solutions, corresponding to apex $Q_1$.} Apex
$Q_1=(-4,1)$ is corresponded to reduced equation \eqref{1.10}.

Let's find the reduced solutions
\begin{equation}
\label{1.17}w=c_r z^r,\,\,\, c_r\neq0
\end{equation}
for $\omega (1,r) \in U_1^{(0)}$.

Since $p_1 < 0$ in the cone $U_1^{(0)}$, then $\omega=-1, \,\,\,
z\rightarrow 0$ and the expansions are the ascending power series of
$z$. The dimension of the bound  $d=0$, therefor
\begin{equation}
\label{1.18}g(z,w)=w^4 \,w^{-1} \,w_{zzzz}
\end{equation}

We get the characteristic polynomial
\begin{equation}
\label{1.19}\chi(r)\stackrel{def}{=} g(z,z^r) = r(r-1)(r-2)(r-3)
\end{equation}
Its roots are
\begin{equation}
\label{1.20}r_1=0,\,\,\, r_2=1,\,\,\, r_3=2,\,\,\, r_3=3
\end{equation}

Let's explore all these roots.

The root $r_1=0$ is corresponded to vector $R=(1,0)$ and vector
$\omega R\in U_1^{(0)}$.

We obtain the family $\mathcal{F}_1^{(1)}\,1$ of reduced solutions
$y=c_0$, where   $c_0 \neq 0$ is arbitrary constant and $\omega =
-1$. The first variation of equation \eqref{1.10}
\begin{equation}
\label{1.21}\frac{\delta\hat{\emph{f}}_1^{(0)}}{\delta w} =
\frac{d^4}{dz^4}
\end{equation}
gives operator
\begin{equation}
\label{1.22}\mathcal{L}(z)=\frac{d^4}{dz^4} \neq 0
\end{equation}
Its characteristic polynomial is
\begin{equation}
\label{1.23}\nu(k) =z^{4-k} \mathcal{L}(z)\, z^k =k(k-1)(k-2)(k-3)
\end{equation}
Equation
\begin{equation}
\label{1.24}\nu(k)=0
\end{equation}
has four roots
\begin{equation}
\label{1.25}k_1=0,\,\,\, k_2=1,\,\,\, k_3=2,\,\,\, k_4=3
\end{equation}

As long as $\omega=-1$ and $r=0$, then the cone of the problem is
\begin{equation}
\label{1.26}\mathcal{K}=\{k>0\}
\end{equation}
It contains the critical numbers $k_2=1,\,\,\, k_3=2$ and $k_3=3$.
Expansions for  the solutions, corresponding to reduced solution
\eqref{1.17} can be presented in the form
\begin{equation}
\label{1.27}w=c_0 + c_1 z + c_2z^2 + c_3z^3 + \sum^{\infty}_{k=4}
c_k z^k
\end{equation}
where all the coefficients are constants, $c_0\neq 0,\,\,\,
c_1,\,\,\,c_2$, $c_3$ are arbitrary ones and  $c_k \,\, (k\geq 4)$
are uniquely defined. Denote this family as $\emph{G}_1^{(0)}1$.
Expansion \eqref{1.27} with taking into account eight terms is
\begin{equation*}\begin{gathered}
\label{eq1.28}w(z)=c_{{0}}+c_{{1}}\,z+c_{{2}}\,{z}^{2}+c_{{3}}\,{z}^{3}+\left
(\frac56\,c_{{0}}c_{{2}}-{\frac {5} {12}}\,{c_{{0}}}^{3}+{\frac
{5}{24}}\,{c_{{1}}}^{2 }\right ){z}^{4}+
\\+\left ({\frac
{1}{120}}\,
\alpha-\frac14\,c_{{1}}{c_{{0}}}^{2}+\frac12\,c_{{0}}c_{{3}}+\frac13\,c_{{1}}c_{{2}}\right
){z}^{5}+
\\+\left ({ \frac {7}{36}}\,c_{{2}}{c_{{0}}}^{2}-{\frac
{5}{36}}\,{c_{{0}}}^{4}-{ \frac
{1}{72}}\,c_{{0}}{c_{{1}}}^{2}+\frac19\,{c_{{2}}}^{2}+\frac14\,c_{{1}}c_
{{3}}\right ){z}^{6}+
\\+\left (\frac{1}{36}\,{c_{{1}}}^{3}+{\frac
{1}{504}}\,c_{{0
}}\alpha+\frac{1}{12}\,c_{{3}}{c_{{0}}}^{2}+\frac16\,c_{{0}}c_{{1}}c_{{2}}-{\frac
{5}{36}}\,c_{{1}}{c_{{0}}}^{3}+\frac16\,c_{{2}}c_{{3}}\right
){z}^{7}+...
\end{gathered}\end{equation*}

Let's explore root $r_2=1$. The cone of the problem is
$\mathcal{K}=\{k>1\}$. It contains the critical numbers
$k_2=2,\,\,\, k_3=3$. The expansion of solution, corresponding to
the reduced solution
\begin{equation*}\mathcal{F}_1^{(1)}2: \,\,\, w= c_1\,z\end{equation*}
can be written as
\begin{equation}
\label{1.29}w(z)=c_1z + c_2z^2 + c_3z^3 + \sum^{\infty}_{k=4} c_kz^k
\end{equation}
where $c_1\neq 0,\,\,\, c_2$ and $c_3$ are the arbitrary constants.
Denote this family as $\emph{G}_1^{(0)}2$. The expansion of
solutions \eqref{1.29} with taking into account seven terms is
\begin{equation*}\begin{gathered}
\label{1.29a}w(z)=c_{{1}}z+c_{{2}}{z}^{2}+c_{{3}}{z}^{3}+{\frac
{5}{24}}\,{c_{{1}}} ^{2}{z}^{4}+\left ({\frac
{1}{120}}\,+\frac13\,c_{{1}}c_{{2}}\right ){z}^{5}+
\\+\left
(\frac14\,c_{{1}}c_{{3}}+\frac19\,{c_{{2}}}^{2}\right ){z}^{6}+
\left (\frac16\,c_{{2}}c_{{3}}+\frac{1}{36}\,{c_{{1}}}^{3}\right
){z}^{7}+...
\end{gathered}\end{equation*}

For root $r_2=2$ the cone of the problem is $\mathcal{K}=\{k>2\}$.
The critical number is $k_3=3$. The expansion of the solutions,
corresponding to the reduced solution
\begin{equation*}\mathcal{F}_1^{(1)}3: \,\,\, w= c_2\,z^{2}\end{equation*}
takes the form
\begin{equation}
\label{1.30}w=c_2z^2 +c_3z^3 + \sum^{\infty}_{k=4} c_k\,z^k
\end{equation}
Denote this family as $\emph{G}_1^{(0)}3$. Expansion \eqref{1.30}
with taking into account eight terms is
\begin{equation}\begin{gathered}
\label{eq1.30a}w(z)=c_{{2}}{z}^{2}+c_{{3}}{z}^{3}+{\frac
{1}{120}}\,{z}^{5}+\frac19\,{c_{{2}}}^{2}{z}^{6}+\frac16\,c_{{2}}c_{{3}}{z}^{7}+\frac{1}{16}\,{c_{{3}}}^{2
}{z}^{8}+
\\+{\frac {1}{1134}}\,c_{{2}}\,{z}^{9}+\left ({\frac {5}{
648}}\,{c_{{2}}}^{3}+{\frac {41}{60480}}\,c_{{3}}\right
){z}^{10}+...
\end{gathered}\end{equation}

For root $r_3=3$ the cone of the problem is $\mathcal{K}=\{k>3\}$.
There is no critical number here. The expansion of solutions,
corresponding to the reduced solution
\begin{equation*}\mathcal{F}_1^{(1)}4: \,\,\, w= c_3\,z^{3}\end{equation*}
takes the form
\begin{equation}
\label{1.31}w(z)=c_3z^3 + \sum^{\infty}_{k=4} c_k\,z^k
\end{equation}
Denote this family as $\emph{G}_1^{(0)}4$. The expansion
\eqref{1.31} with taking into account four terms is
\begin{equation}\begin{gathered}
\label{eq1.31a}w(z)=c_{{3}}{z}^{3}+{\frac
{1}{120}}\,{z}^{5}+\frac{1}{16}\,{c_{{3}}}^ {2}{z}^{8}+{\frac
{41}{60480}}\,c_{{3}}\,{z}^{10} + \ldots
\end{gathered}\end{equation}

The expansions of solutions converge for sufficiently small $|z|$.
The existence and analyticity of expansions \eqref{1.27},
\eqref{1.29}, \eqref{1.30} and \eqref{1.31} follow from Cauchy
theorem.

\section{\!\!\!\!\!\!.\,\, Solutions, corresponding to edge $\Gamma^{(1)}_1$.} Edge $\Gamma^{(1)}_1$
is conformed by the reduced equation
\begin{equation}
\label{1.32}\hat{f}_1^{(1)} (z,y)\stackrel{def}{=}y_{zzzz} - z=0
\end{equation}
Normal cone is
\begin{equation}
\label{1.33}U^{(1)}_1 =\{-\mu(1,5),\,\,\mu>0\}
\end{equation}
Therefor $\omega=-1$, i.e. $z\rightarrow 0$ and $r=5$. Power
solutions are found in the form
\begin{equation*}
\label{1.34} w=c_5z^5
\end{equation*}
For $c_5$ we have
\begin{equation}
\label{1.35}c_5=\frac {1}{120}
\end{equation}
The only power solution is
\begin{equation}
\label{1.36}\mathcal{F}_2^{(1)}1: \,\,\,w=\frac{z^5}{120}
\end{equation}
Compute the critical numbers. The first variation of \eqref{1.13} is
\begin{equation}
\label{1.37}\frac{\delta \hat{f}^{(1)}_1}{\delta w}
=\frac{d^4}{dz^4}
\end{equation}
We get the proper numbers
\begin{equation}
\label{1.38}k_1=0,\,\,\, k_2=1,\,\,\, k_3=2,\,\,\, k_4=3
\end{equation}
The cone of the problem
\begin{equation*}
\mathcal{K}=\{k>5\}
\end{equation*}
doesn't consist them. Solution \eqref{1.36} is corresponded to two
vector indexes $\tilde{Q}_1=(0,1),\,\,\, \tilde{Q}_2= (5,0)$. There
difference  $B=\tilde{Q}_1 - \tilde{Q}_2 =(-5,\,1)$ equals to vector
$Q_1-Q_2$. So solution \eqref{1.36} is conformed to lattice
$\textbf{Z}$, which consists of points $Q=(q_1,\, q_2)=k(-3,\,2)
+m(-5,\,1)=(-3k-5l,\,2k+l)$, where $k$  and $l$ are whole numbers.
Points belong to line $q_2=-1$, if $l=-1-2k$. In this case
$q_1=5+7k$. As long as the cone of the problem here is
$\mathcal{K}=\{k>5\}$, the set of the carrier of solution expansion
$\textbf{K}$ takes the form
\begin{equation}
\label{1.39}\textbf{K}=\{5+7n,\,\,\,n\in \mathbb{N}\}
\end{equation}
Then the expansion of solution can be written as
\begin{equation}
\label{1.40}w(z)=z^5\left(\frac{1}{120}+\sum^{\infty}_{m=1}
c_{5+7m}\,z^{7m}\right)
\end{equation}
Expansion \eqref{1.40} with taking into account three terms takes
the form
\begin{equation}
\label{1.41}w(z)=\frac{z^5}{120}\left(1+\frac{13\,z^7}{57024}+\frac{2851\,z^{14}}{79569008640}+...\right)
\end{equation}
Equation \eqref{1.32} doesn't have exponential additions and
non-power asymptotic forms.

\section{\!\!\!\!\!\!.\,\, Solutions, corresponding to edge $\Gamma^{(1)}_2$.} Edge $\Gamma^{(1)}_2$
is corresponded to the reduced equation
\begin{equation}
\label{1.42}\hat{f}^{(1)}_2 (z,w) \stackrel{def}{=}w_{zzzz}
-10\,w\,w_{zz} -5\,w^2+10\,w^3=0\end{equation}

The normal cone is
\begin{equation}
\label{1.43}U^{(1)}_2=\{-\mu(1,-2),\,\mu>0\}
\end{equation}
Therefor $\omega=-1$, i.e. $z\rightarrow 0$ and $r=-2$. Hence the
solution of equation \eqref{1.42} we can find in the form
\begin{equation}
\label{1.44}w=c_{-2} z^{-2}
\end{equation}
For $c_{-2}$ we have the determining equation
\begin{equation}
\label{1.45}c_{-2}^2 -8\,c_{-2} +12 =0
\end{equation}
Consequently we get
\begin{equation}
\label{1.46}c_{-2}^{(1)} =2,\,\,\quad\, c_{-2}^{(2)}=6
\end{equation}
The reduced solutions are
\begin{equation}
\label{1.47}\mathcal{F}_2^{(1)}1: \,\,\, w=2z^{-2}
\end{equation}
\begin{equation}
\label{1.48}\mathcal{F}_2^{(1)}2: \,\,\, w=6z^{-2}
\end{equation}
Let's compute the corresponding critical numbers. The first
variation is
\begin{equation}
\label{1.49}\frac{\delta f_2^{(1)}}{\delta w}=\frac{d^4}{dz^4}
-10w_{zz} -10w\frac{d^2}{dz^2} -10\,w_z \frac d{dz} + 30\,w^2
\end{equation}
Applied to  solution \eqref{1.47}, it produces operator
\begin{equation}
\label{1.50}\mathcal{L}^{(1)}(z) =\frac{d^4}{dz^4}-\frac{20}{z^2}
\frac{d^2}{dz^2} + \frac{40}{z^3} \frac{d}{dz}
\end{equation}
which is corresponded by the characteristic polynomial
\begin{equation}
\label{1.51}\nu(k ) =k^4 -6k^3 -9k^2 + 54k
\end{equation}
Equation
\begin{equation}
\label{1.52}\nu(k)=0
\end{equation}
has the roots
\begin{equation}
\label{1.53}k_1=-3,\,\,\, k_2=0,\,\,\, k_3=3,\,\,\, k_4=6
\end{equation}
With reference to solution \eqref{1.48} variation \eqref{1.49} gives
operator
\begin{equation}
\label{1.54}\mathcal{L}^{(2)}(z) =\frac{d^4}{dz^4} + \frac{60}{z^2}
\frac{d^2}{dz^2} -\frac{120}{z^3} \frac{d}{dz} + \frac{720}{z^4}
\end{equation}
which is corresponded by the characteristic polynomial
\begin{equation}
\label{1.55}\nu(k)=k^4 -6k^3 -49k^2 +174k +720
\end{equation}
with roots
\begin{equation}
\label{1.56}k_1=-5,\,\,\, k_2=-3,\,\,\, k_3=6,\,\,\, k_4=8
\end{equation}
The cone of the problem here is
\begin{equation}
\label{1.57} \mathcal{K}=\{k>-2\}
\end{equation}
Therefor for the reduced solution \eqref{1.47} three critical
numbers belong to the cone, and there are two critical numbers for
the reduced solution \eqref{1.48} in the cone of the problem.

The set of the carriers of the solution expansions $\textbf{K}$ can
be written as
\begin{equation}
\label{1.57a}\textbf{K}=\{-2+7n,\,\,\,n\in \mathbb{N}\}
\end{equation}
Sets $\textbf{K}(0)$, $\textbf{K}(0,3)$ and $\textbf{K}(0,3,6)$ are
\begin{equation} \begin{gathered}
\label{1.57b}\textbf{K}(0)=\{-2+7n+2m,\,\,n,m\in
\mathbb{N},\,\,n+m\geq\,0\}=\\
=\{-2,0,4,6,5,7,8,...\}
\end{gathered}\end{equation}
\begin{equation}\begin{gathered}
\label{1.57c1}\textbf{K}(0,3)=\{-2+7n+2m+5k,\,n,m,k\in
\mathbb{N},\,\,m+n+k\geq\,0\}=\\
=\{-2,0,2,3,4,5,6,7,8,...\}
\end{gathered} \end{equation}
\begin{equation}\begin{gathered}
\label{1.57c2}\textbf{K}(0,3,6)=\{-2+7n+2m+5k+8l,\,n,m,k,l \in
\mathbb{N}
,\,\,m+n+k+l\geq\,0\}=\\
=\{-2,0,2,3,4,5,6,7,8,...\}
\end{gathered}\end{equation}
In this case the expansion for the solution of equation can be
represented as
\begin{equation}\begin{gathered}
\label{eq1.57d}w(z)=\frac{2}{z^2}+ \sum^{}_{n,m,k,l}
c_{5+7n+2m+2k+8l}\,z^{5+7n+2m+2k+8l}
\end{gathered}\end{equation}
Denote this family as $G_{2}^{1}1$. The critical number $0$ doesn't
belong to set $\textbf{K}$, so the compatibility condition for $c_0$
holds automatically and  $c_0$ is the arbitrary constant. The
critical number $3$ also doesn't belong to sets $\textbf{K}$ and
$\textbf{K}(0)$, therefor the compatibility condition for $c_3$
holds too and $c_3$ is the arbitrary constant. But critical number
$6$ is a member of $\textbf{K}(0)$ and $\textbf{K}(0,6)$, so it is
necessary  to verify that the compatibility condition for $c_6$
holds and that $c_6$ is the arbitrary constant. The calculation
shows that in this situation the condition holds and $c_6$ is the
arbitrary constant too. The three-parameter power expansion of
solutions, corresponding to the reduced solution  \eqref{1.47} takes
the form
\begin{equation}\begin{gathered}
\label{1.58}
w(z)=\frac{2}{{z}^{2}}+c_{{0}}-\frac32\,{c_{{0}}}^{2}{z}^{2}+c_{{3}}{z}^{3}-\frac52
\,{c_{{0}}}^{3}{z}^{4}+\left (\frac34\,c_{{0}}c_{{3}}-{\frac
{1}{80}}\right ){z}^{5}+c_{{6}}{z}^{6}-
\\-{\frac {1}{280}}\,c_{{0}}\, {z}^{7}+\left
({\frac {153}{352}}\,{c_{{0}}}^{5}+{\frac {9}{44}}\,c_{{0
}}c_{{6}}+{\frac {9}{176}}\,{c_{{3}}}^{2}\right ){z}^{8}+\left
({\frac {19}{12096}}\,{c_{{0}}}^{2 }-{ \frac
{5}{16}}\,{c_{{0}}}^{3}c_{{3}}\right ){z}^{9}+
\\+\left ({\frac
{25}{104}}\,{c_{{0}}}^{6}-{\frac {29}{29120}}\,c_{{3}}-{\frac
{3}{26}}\,{c_{{0}}}^{2}c_{{6}}+{ \frac
{3}{52}}\,c_{{0}}{c_{{3}}}^{2}\right ){z}^{10}+...
\end{gathered}
\end{equation}

The carrier of power expansion, corresponding to  reduced solution
\eqref{1.48}, is formed by the sets
\begin{equation}\begin{gathered}
\label{1.58a}\textbf{K}(6)=\{-2+7n+8m,\,n,m\in
\mathbb{N},\,\,m+n\geq\,0\}=\\
=\{-2,5,6,12,14,20,21,22,27,28,29,30,34,35,36,37,38,41,...\}
\end{gathered}\end{equation}
\begin{equation}\begin{gathered}
\label{1.58b}\textbf{K}(6,8)=\{-2+7n+8m+10k,\,n,m,k\in
\mathbb{N},\,\,m+n+k\geq\,0\}=\\
=\{-2,5,6,8,12,13,14,15,16,18,19,20,21,...\}
\end{gathered}\end{equation}
The expansion for solution of equation can be written as
\begin{equation}\begin{gathered}
\label{eq1.58c}w(z)=\frac{6}{z^2}+ \sum^{}_{n,m,k}
c_{5+7n+8m+10k}\,z^{5+7n+8m+10k}
\end{gathered}\end{equation}
Denote this family as $G_{2}^{1}2$. The critical numbers  6 and 8
don't belong to the set $\textbf{K}$ and the number 8 doesn't belong
to the set $\textbf{K}(6)$. For numbers 6 and 8 the compatibility
conditions holds automatically, therefor coefficients  $c_6$ and
$c_8$ are the arbitrary constants. The two-parameter expansion of
solution, corresponding to the reduced solution \eqref{1.48}, is
\begin{equation}\begin{gathered}
\label{1.59}w(z)=\frac{6}{{z}^{2}}+{\frac
{1}{240}}\,{z}^{5}+c_{{6}}{z}^{6}+c_{{ 8}}{z}^{8}+{\frac
{29}{70502400}}\,{z}^{12}+
\\+{\frac {11}{ 60480}}\,c_{{6}}{z}^{13}+{\frac
{25}{1292}}\,{c_{{6}}}^{2}{z}^ {14}+{\frac
{1}{6804}}\,c_{{8}}{z}^{15}+...
\end{gathered}\end{equation}
According to \cite{Bruno02}, the expansions of solutions
\eqref{1.58} and \eqref{1.59} don't have power and exponential
additions.

\section{\!\!\!\!\!\!.\,\, Solutions, corresponding to edge $\Gamma^{(1)}_3$.} Edge
$\Gamma^{(1)}_3$ is corresponded by the reduced equation
\begin{equation}
\label{1.60}\hat{f}_3^{(1)} (z,w) \stackrel{def}{=} 10\,w^3 - z=0
\end{equation}
It has three power solutions
\begin{equation}
\label{1.61}\mathcal{F}_3^{(1)}1:\,\,\,\, w=\varphi^{(1)}(z)=
c_{1/3}^{(1)}\,z^{1/3},\,\quad\,c_{1/3}^{(1)}=\left(\frac{1}{10}\right)^{1/3}
\end{equation}
\begin{equation}
\label{1.62}\mathcal{F}_3^{(1)}2:\,\,\,\,
w=\varphi^{(2)}(z)=c_{1/3}^{(2)}\,
z^{1/3},\,\quad\,c_{1/3}^{(2)}\,=\left(\frac 12+
i\sqrt{3}\right)\left(\frac{1}{10}\right)^{1/3}
\end{equation}
\begin{equation}
\label{1.63}\mathcal{F}_3^{(1)}3:\,\,\,\,w=\varphi^{(3)}(z)=c_{1/3}^{(3)}\,z^{1/3},\,\quad\,c_{1/3}^{(3)}\,=\left(\frac
12- i\sqrt{3}\right)\left(\frac {1}{10}\right)^{1/3}
\end{equation}
The shifted carrier of reduced solutions \eqref{1.61} --
\eqref{1.63} gives a vector
\begin{equation}
\label{1.64}B=\left(\frac13,-1\right)
\end{equation}
which equals a third of vector $Q_2-Q_1$. Therefor we explore the
lattice, generated by vectors $Q_3-Q_1$ and $B$. We have
$Q=(q_1,q_2) =k(-5,\,1) +l(-3,\,2)+ m\left(\frac13,\,-1\right)
=\left(2k+\frac l3,\,\,k-l \right)$, where $k$, $l$  and $m$ are the
whole numbers. At the line $q_2=-1$ we have $k+2\,l-l-m=-1$,
wherefrom $m=k+2l+1$ and $q_1=\frac{(1+7l-14k)}{3}$. And so the
carrier of solution is
\begin{equation}
\label{1.65}\mathbf{K}=\left\{k=\frac{1-7n}{3},\,\,\, n\in
\mathbb{N}\right\}
\end{equation}
and the expansions of solutions take the form
\begin{equation}
\label{1.66}G_3^{(1)} l:\,\,\,\, w=\varphi^{(l)}(z)=c^{(l)}_{1/3}
z^{1/3} + \sum^{\infty}_{n=1} c^{(l)}_{(1-7n)/3}\, z^{{(1-7n)}/{3}}
\end{equation}
Here $c^{(l)}_{1/3}$ can be found from reduced solutions
\eqref{1.61} -- \eqref{1.63}, coefficients $c^{(l)}_{(1-7n)/3}$ are
computed sequentially. The calculating of the coefficient $c_{-2}$
gives the result $c_{-2}=-\frac{1}{18}$. The expansion of solution
with taking into account five terms is
\begin{equation}
\begin{gathered}
\label{1.66a}\varphi^{(l)}(z)=c_{{1/3}}\,{z^{1/3}}-\frac1{18}\,{z}^{-2}-{\frac
{7}{108}}\,{\frac {1}{c_{{ 1/3}}}}{z}^{-13/3}\,-\\
\\
-{\frac {4199}{17496}}\,{c_{{1/3}}}^{-2}\,{z}^{-{{20}/{3}}}-{\frac
{28006583}{23514624}}\,{\frac {1}{{c_{{1/3}}}^{3}}}{z}^ {-9}\,+ ...
\end{gathered}
\end{equation}
The obtained expansions seem to be divergent ones.

\section{\!\!\!\!\!\!.\,\, Exponential additions of the first level.}

Let's find the exponential additions to solutions
\eqref{1.61}-\eqref{1.63}. We look for the solutions in the form

\begin{equation*} \label{1.67}w=\varphi^{(l)}(z) + u^{(l)},\,\,\,
l=1,2,3
\end{equation*}
The reduced equation for the addition  $u^{(l)}$ is
\begin{equation}
\label{2.59}M_{l}^{(1)}(z) u^{(l)}=0
\end{equation}
where $M_{l}^{(1)}(z)$ is the first variation at the solution
$w=\varphi^{(l)}(z)$. As long as
\begin{equation}
\label{2.60}\frac{\delta f}{\delta w} =\frac{d^4}{dz^4} - 10 w_{zz}
-10w\frac{d^2}{dz^2} -10w_z \frac{d}{dz} +30w^2
\end{equation}
then
\begin{equation}
\label{2.61}M_{l}^{(1)}(z) =\frac{d^4}{dz^4} -10\varphi^{(l)}_{zz}
-10 \varphi^{(l)}\frac{d^2}{dz^2} -10\varphi_z^{(l)} \frac d{dz} +
30{\varphi^{(l)}}^2
\end{equation}
Equation \eqref{2.59} takes the form
\begin{equation}
\begin{gathered} \label{2.62}\frac{d^4u^{(l)}}{dz^4} -10 \varphi^{(l)}_{zz} u^{(l)} -10
\varphi{(l)} \frac{d^2u^{(l)}}{dz^2} -10 \varphi^{(l)} _z
\frac{du^{(l)}}{dz}
+30{\varphi^{(l)}}^2 u^{(l)} =0,\,\,\, \\
 l=1,2,3
\end{gathered}
\end{equation}
Suppose that
\begin{equation}
\label{2.63}\zeta^{(l)}=\frac{d \ln u^{(l)}}{dz}
\end{equation}
then from \eqref{2.63} we have
\begin{equation*} \label{}\frac{du^{(l)}}{dz} =\zeta^{(l)} u^{(l)},
\,\,\quad\, \frac{d^2u^{(l)}}{dz^2 }=\zeta _z^{(l)} u^{(l)} + {\zeta
^{(l)}}^2 u^{(l)}
\end{equation*}
\begin{equation*}
\label{}\frac{d^3u^{(l)}}{dz^3} =\zeta^{(l)}_{zz} u^{(l)}+ 3 \zeta
^{(l)} \zeta_z^{(l)} u^{(l)} + {\zeta^{(l)}}^3 u^{(l)}
\end{equation*}
\begin{equation*}
\label{}\frac{d^4u^{(l)}}{dz^4} =\zeta^{(l)}_{zzz} u^{(l)}+ 4 \zeta
^{(l)}\zeta^{(l)}_{zz} u^{(l)} + 3 {\zeta_z^{(l)}}^2  u^{(l)} + 6
{\zeta^{(l)}}^2 \zeta _z^{(l)} u^{(l)} + {\zeta ^{(l)}}^4 u^{(l)}
\end{equation*}
By substituting the derivatives
\begin{equation*}
\label{}\frac{du^{(l)}}{dz},\,\quad \, \frac{d^2 u^{(l)}}{dz^2} ,
\,\quad \,\frac{d^4u^{(l)}}{dz^4}
\end{equation*}
into the equation \eqref{2.62} we get the reduced equation in the
form
\begin{equation}
\begin{gathered}
\label{6.13}u^{(l)} \left[\zeta^{(l)}_{zzz} +
4\zeta^{(l)}\zeta^{(l)}_{zz}
+3{\zeta^{(l)}_z}^2 +6 {\zeta^{(l)}}^2 \zeta^{(l)}_z + \right. \\
+ \left. {\zeta^{(l)}}^4 -10\varphi_{zz}^{(l)} -10\varphi^{(l)}
\zeta^{(l)}_z -10\varphi^{(l)} {\zeta^{(l)}}^2 -10\varphi_z^{(l)}
\zeta^{(l)}+30 {\varphi^{(l)}}^2\right]=0
\end{gathered}
\end{equation}

Let's find the power expansions for solutions of equation
\eqref{6.13}. The carrier of equation \eqref{6.13} consists of
points
\begin{equation}
\begin{gathered}
\label{2.65}Q_1 =(-3,1),\,\,\, Q_2=(-2,2),\,\,\, Q_3=(-1,3), \\
Q_4=(0,4), \,\,\, Q_{5} =\left(\frac13,2\right),\,\,\, Q_{6}
=\left(\frac23,0\right),
Q_{7}=\left(-\frac23,1\right),\\
Q_{8}=\left(-\frac53,0\right),\,\,
Q_{8+n}=\left(-\frac{5+7n}3,0\right),\,\,
Q_{9+m}=\left(-\frac{2+7m}3,1\right),\\
Q_{10+k}=\left(\frac{1-7k}3,2\right),\,\,
Q_{11+k}=\left(\frac{12-14l}3,0\right),\,\,\, m,n,k,l \in \mathbb{N}
\end{gathered}
\end{equation}
The closing of convex hull of points of the carrier of equation
\eqref{6.13} is the strip. It is represented at fig. 3.

%Тут надо вставить зашибенную картинку

\begin{figure}[h] % было [p]
 \centerline{\epsfig{file=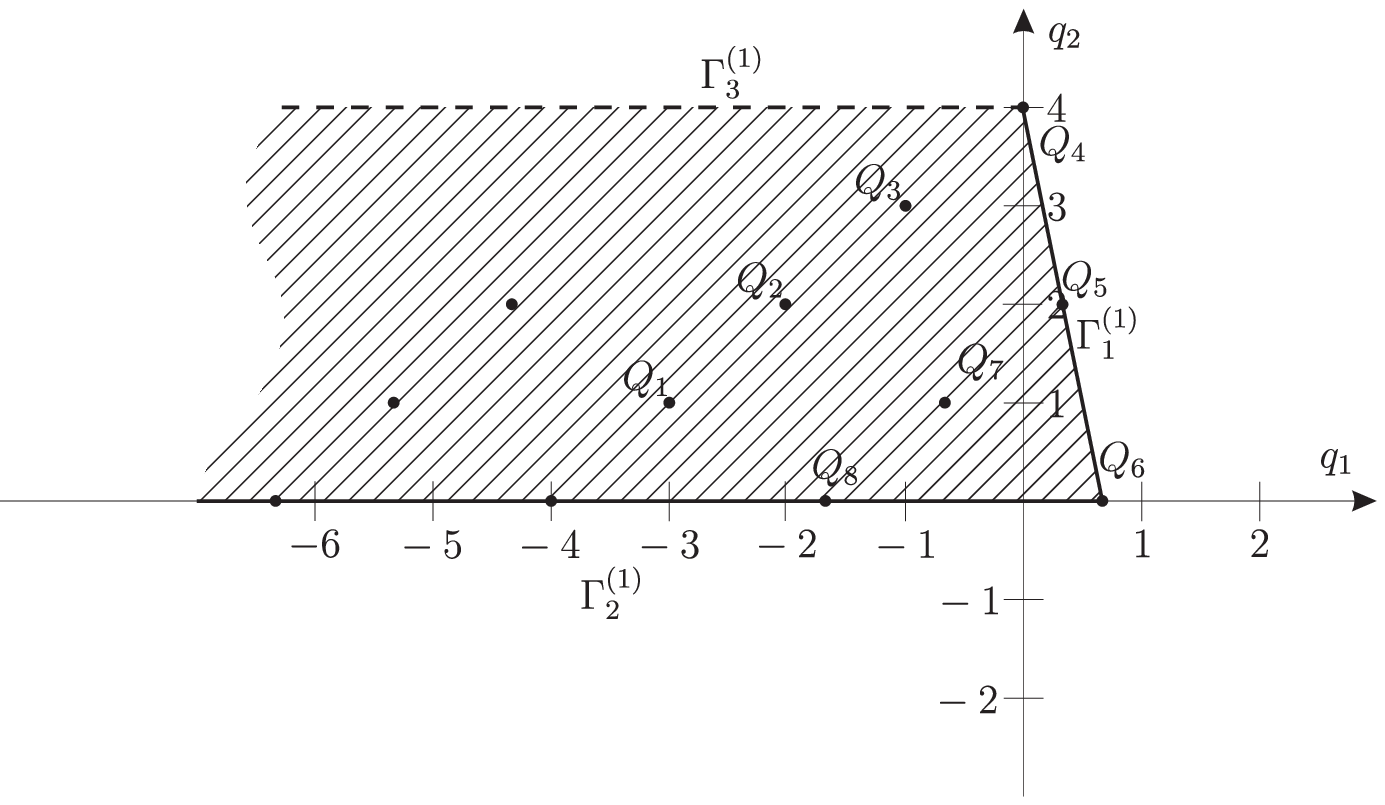,width=120mm}}
 \caption{}\label{fig:z_post_1}
\end{figure}

The periphery of the strip contains edges
$\Gamma^{(1)}_j\,\,(j=1,2,3)$ with normal vectors $N_1=(6,1),\,\,
N_2=(0,-1),\,\, N_3=(0,1)$. It should take up edge $\Gamma^{(1)}_1$
only . This edge is corresponded by the reduced equation
\begin{equation}
\label{6.15}h_1^{(1)}(z,\zeta) \stackrel{def}{=} \zeta^4
-10\varphi^{(l)} \zeta^2 +30{\varphi^{(l)}}^2=0
\end{equation}
Wherefrom we have
\begin{equation}
\label{6.16}\zeta^2 =\left(5+(-1)^{m-1} i \sqrt{5}\right)
\varphi^{(l)},\,\,\, m=1,2
\end{equation}
We obtain twelve solutions of equation  \eqref{6.15}
\begin{equation}\begin{gathered}
\label{6.17}\zeta^{(l,m,k)} =g_{1/6}^{(l,m,k)}z^{1/6},\,\,\,
l=1,2,3;\,\, m,k=1,2 \end{gathered}\end{equation} where
\begin{equation}\begin{gathered}
\label{6.18}g_{1/6}^{(1,m,k)}=\left(\frac1{10}\right) ^{1/3}
(-1)^{k-1} \left(5+(-1)^{m-1} i\sqrt{5}\right)^{1/2},\\
\,\,\,m,k=1,2
\end{gathered}\end{equation}
\begin{equation}\begin{gathered}
\label{6.19}g_{1/6}^{(2,m,k)}=\left(\frac12+
i\sqrt{3}\right)\left(\frac1{10}\right) ^{1/3} (-1)^{k-1}
\left(5+(-1)^{m-1} i\sqrt{5}\right)^{1/2},\\
\,\,\, m,k=1,2
\end{gathered}\end{equation}
\begin{equation}\begin{gathered}
\label{6.20}g_{1/6}^{(3,m,k)}=\left(\frac12-
i\sqrt{3}\right)\left(\frac1{10}\right) ^{1/3} (-1)^{k-1}
\left(5+(-1)^{m-1} i\sqrt{5}\right)^{1/2},\\
\,\,\, m,k=1,2
\end{gathered}\end{equation}

The reduced equation is algebraic one, so it have no critical
numbers. Let's compute the carrier of the expansion for solution of
equation \eqref{6.13}. The shifted carrier of equation \eqref{6.13}
is contained in a lattice, generated by vectors
$B_1=\left(\frac73,0\right),\,\, B_2=(1,1)$. The shifted carrier of
solutions \eqref{6.17} gives rise to vector $B_3=\left(-\frac16,1
\right)$. The difference $B_2-B_3=\left(\frac76,0\right)=\frac12 B_2
\stackrel{def}{=} B_4$. Therefor, vectors $B_1,B_2 $
 and $B_3$ generate the same lattice as vectors $B_2,B_4$. Points of this lattice can be written as
\begin{equation*}
Q=(q_1,q_2)=k(1,1)
+m\left(\frac76,0\right)=\left(k+\frac{7m}6,k\right)
\end{equation*}
At the line $q_2=-1$ we have $k=-1$, and so $q_1=-1+\frac{7m}6$. As
long as the cone of the problem here is
$\mathcal{K}=\left\{k<\frac16\right\}$, then the set of the carriers
of expansions $\mathbf{K}$ is
\begin{equation}
\label{6.21}\mathbf{K}=\left\{\frac{1-7n}6,n\in\mathbb{N}\right\}
\end{equation}
The expansion for solution of equation  \eqref{6.13} takes the form
\begin{equation}\begin{gathered}
\label{6.22.1}\zeta^{(l,m,k)}=g^{(l,m,k)}_{1/6} z^{1/6}
+\sum_{n}g^{(l,m,k)}_{(1-7n)/6}\,z^{(1-7n)/6},\,\,\,\\
 l=1,2,3;\,\,\quad\,
m=1,2;\,\,\quad\, k=1,2 \end{gathered}\end{equation} Coefficients
$g_{1/6}^{(l,m,k)}$ are determined by expressions \eqref{6.18},
\eqref{6.19} and \eqref{6.20}. Coefficient $g^{(l,m,k)}_{-1}$ takes
on a value
\begin{equation}
\label{6.22a}
g_{-1}^{(l,m,k)}=-\frac1{4}\\
\end{equation}
The expansion of solution with taking into account four terms takes
the form
\begin{equation}
\begin{gathered}
\label{6.22b}\zeta^{(l,m,k)}=g_{{1/6}}{z^{1/6}}-\frac14\,{z}^{-1}
-{\frac {7}{288}}\, \frac{\left(30\,{g_{{1/6}}}^{2}-7\,{10} ^{2/3}
\right)
}{{g_{{1/6}}} \left( 2\,{g_{{1/6}}}^{2}-{10}^{2/3} \right)   }{z}^{-13/6}\,-\\ \\
-{\frac {49}{ 1728}}\,{\frac
{30\,{g_{{1/6}}}^{4}-6\,{10}^{2/3}{g_{{1/6}}}^{2}+35\, \sqrt [3]{10}
}{{g_{{1/6}}}^{2} \left( 5\,\sqrt [3]{10}-2\,{10}^{2/3}\, {g
_{{1/6}}}^{2}+2\,{g_{{1/6}}}^{4} \right) }}{z}^{-10/3}\,+ ...
\end{gathered}
\end{equation}

%\begin{equation}
%\begin{gathered}
%\label{6.22b}\zeta^{(l,m,k)}=g_{{1/6}}\sqrt [6]{z}-\frac14\,{z}^{-1}-\\ \\
%-{\frac {7}{288}}\, \left( -7\,{10} ^{2/3}+30\,{g_{{1/6}}}^{2} \right) {g_{{1/6}}}^{-1}
%\left( -{10}^{2/3} +2\,{g_{{1/6}}}^{2} \right) ^{-1}{z}^{-{\frac {13}{6}}}-\\ \\
%-{\frac {49}{ 1728}}\,{\frac {-6\,{10}^{2/3}{g_{{1/6}}}^{2}+30\,{g_{{1/6}}}^{4}+35\,
%\sqrt [3]{10}}{{g_{{1/6}}}^{2} \left( 5\,\sqrt [3]{10}-2\,{10}^{2/3}{g
%_{{1/6}}}^{2}+2\,{g_{{1/6}}}^{4} \right) {z}^{10/3}}}-\\ \\
%-{\frac {35}{ 165888}}\, \left( {145015\,{10}^{2/3}{g_{{1/6}}}^{2}+279140\,{g_{{1/6}}
%}^{4}-305364\,{g_{{1/6}}}^{6}\sqrt [3]{10}+ }\right. \\
%\left. {+67074\,{g_{{1/6}}}^{8}{10}^ {2/3}-66036\,{g_{{1/6}}}^{10}-214375\,\sqrt
%[3]{10}+3072\,\sqrt [3]{10
%}{g_{{1/6}}}^{12}}\right) /\\
%\\/
%\left({g_{{1/6}}}^{3} \left( 125\,{10}^{2/3}{g_{{1/6}}}^{
%2}-500\,{g_{{1/6}}}^{4}+100\,{g_{{1/6}}}^{6}\sqrt [3]{10}\right. \right.-\\
%\left. \left.-125\,\sqrt [ 3]{10}+4\,{g_{{1/6}}}^{10}-10\,{g_{{1/6}}}^{8}{10}^{2/3}
%\right) {z}^{ 9/2}\right)
%\end{gathered}
%\end{equation}
In view of \eqref{2.63} we can find the additions $u^{(l,m,k)}(z)$.
We have
\begin{equation*}
\label{6.22_2}u^{(l,m,k)}(z) = C \exp \int \zeta^{(l,m,k)}(z) dz
\end{equation*}

Wherefrom we get
\begin{equation}
\begin{gathered}
\label{6.23}u^{(l,m,k)}(z)=C_1\,z^{-1/4}\, \exp \left[\frac67\,
g^{(l,m,k)}_{1/6}\,z^{7/6} + \sum^{\infty}_{n=2} \frac{6}{7(1-n)}
g^{(l,m,k)}_{(1-7n)/6}
z^{7(1-n)/6}\right]\\
\,\, l=1,2,3;\,\quad\,m=1,2;\,\quad\, k=1,2
\end{gathered}
\end{equation}
Here $C_1$ and farther $C_2$ and  $C_3$ are the arbitrary constants.
Addition $u^{(l,m,k)}(z)$ near $z\rightarrow \infty$ is the
exponentially small one in those sectors of complex plane $z$, where
\begin{equation}
\label{6.24}Re \left[g^{(l,m,k)}_{1/6}\, z^{1/6}\right]<0
\end{equation}
Thus for three expansions $G^{(1)}_{3}l$ we get four one-parameter
family of additions $G_3^{(1)}l G^1_1 mk$, where $m=1,2$ and
$k=1,2$.

\section{\!\!\!\!\!\!.\,\, Exponential additions of the second level.}

Let's find exponential additions of the second level $v^{(p)}$, i.e.
the additions to solutions $u^{(l,m,k)}(z)$. The reduced equation
for addition $v^{(p)}$ is
\begin{equation}
\label{6.27}M_{p}^{(2)} (z) v^{(p)}=0
\end{equation}
where operator $M_{p}^{(2)}$ is the first variation of \eqref{6.13}.
Equation \eqref{6.32} for $v=v^{(p)}$ takes the form
\begin{equation}
\begin{gathered}
\label{6.29}\frac{d^3v}{dz^3} + 4\zeta_{zz} v+ 4\zeta v_{zz} +
6\zeta_z v_z + 12
\zeta\zeta_z v+ \\
+6\zeta^2 v_z + 4\zeta ^3 v -10 \varphi ^{(l)} v_z -20\varphi ^{(l)}
\zeta v - 10\varphi_z^{(l)} v=0
\end{gathered}
\end{equation}
Assumed that
\begin{equation}
\label{6.30}\frac{d \ln v}{dz}=\xi
\end{equation}
we have
\begin{equation}
\label{6.31}\frac{dv}{dz}=\xi v,\,\,\quad\, \frac{d^2v}{dz^2} =\xi_z
v+\xi^2 v,\,\,\quad\, \frac{d^3 v}{dz^3}=\xi_{zz} v + 3\xi\xi_{z}
v+\zeta^3 v
\end{equation}
From \eqref{6.29} we get equation
\begin{equation}\begin{gathered}
\label{6.32}\xi_{zz} + 3\, \xi\,\xi_{z} + \xi^3 + 4\zeta_{zz}
+4\xi_z\,\zeta+4\xi^2\,\zeta+ 6\,\xi\,\zeta_z +12 \zeta\zeta_z
+6\,\xi\,\zeta^2 +
\\+4\,\zeta^3 -10\,\varphi^{(l)}\xi -20\,\zeta\, \varphi^{(l)}
-10\,\varphi_z^{(l)}=0
\end{gathered}\end{equation}
Monomials of equation \eqref{6.32} is corresponded by the points
\begin{equation}
\begin{gathered}
\label{6.33}M_1=(-2,1),\,\,\, M_2=(-1,2),\,\,\, M_3=(0,3),\,\,\,
M_4=\left(-\frac{11}6,\,0\right),\,\,\,\\
M_5=\left(-\frac56,1\right),\,\,\,M_6=\left(\frac16,2\right),\,\,\,M_7=\left(-\frac56,\,1\right),\,\,\,\\
M_8=\left(-\frac23,\,0\right),\,\,\,M_9=\left(\frac13,\,1\right),\,\,\,
M_{10}=\left(\frac12,\,0\right),\,\,\,\\
M_{11}=\left(\frac13,1\right),\,\,\,\,
M_{12}=\left(\frac12,\,0\right),\,\,\,
M_{13}=\left(-\frac23,\,0\right),\,\,\,.\,.\,.
\end{gathered}
\end{equation}
The carrier of the equation  \eqref{6.32} is determined by points of
the set \eqref{6.33}. The convex set forms the strip,which is
similar to the strip, represented at fig. 3. It should examine edge
$\Gamma_1^{(1)}$,  which is passing through points
\begin{equation}
\begin{gathered}
\label{6.34}Q_1=\left(\frac12,0\right),\,\,\, Q
_2=\left(\frac13,1\right),\,\,\, Q_3=\left(0,3\right)
\end{gathered}
\end{equation}

The reduced equation, corresponding to this edge, is
\begin{equation}
\begin{gathered}
\label{6.35}\xi^3 +4\,\xi^{2}\,\zeta
+6\,\xi\,\zeta^2+4\,\zeta^3-20\,\zeta\,\varphi^{(l)}
-10\,\xi\,\varphi^{(l)}=0
\end{gathered}
\end{equation}
The basis of the lattice, corresponding to the carrier of equation
\eqref{6.32} is
\begin{equation*}
\begin{gathered}
\label{6.36}B_1=(1,1),\,\,\,\, B_2=\left(\frac76,0\right)
\end{gathered}
\end{equation*}
The solution of equation \eqref{6.35} takes the form
\begin{equation}
\begin{gathered}
\label{6.37}\xi^{(l,m,k,p)}=r_{1/6}^{(l,m,k,p)}\, z^{1/6},\,\,\,\,
m,k=1,2;\,\,\,\, l=1,2,3;\,\,\,\, p=1,2,3
\end{gathered}
\end{equation}
where $r=r_{1/6}^{(l,m,k,p)},\,\,\,p=1,2,3$  are the roots of the
equation
\begin{equation}
\begin{gathered}
\label{6.38}5\,r^3 +4\,r^2\,g_{1/6}^{(l,m,k)}+
\left(6\,g_{1/6}^{(l,m,k)^2} -10\,c_{1/3}^{(l)}\right)\,r+4\, {g_{1/6}^{(l,m,k)}}^{3}\\
-20\,g_{1/6}^{(l,m,k)}\,c_{1/3}^{(l)}=0
\end{gathered}
\end{equation}
Equation \eqref{6.38} has the roots
\begin{equation}
\begin{gathered}
\label{6.38a}r_{1/6}^{(l,m,k,1)}=-2\,g_{1/6}^{(l,m,k)},\,\quad\,
r_{1/6}^{(l,m,k,2)}=-g_{1/6}^{(l,m,k)}+\left({10\,c_{1/3}^{(l)}-g_{1/6}^{(l,m,k)}}\right)^{1/2}\\
r_{1/6}^{(l,m,k,3)}=-g_{1/6}^{(l,m,k)}-\left({10\,c_{1/3}^{(l)}-g_{1/6}^{(l,m,k)}}\right)^{1/2}
\end{gathered}
\end{equation}
The set of carriers of expansions for solution $\mathbf{K}$
coincides with \eqref{6.21}. The expansion of solution for
$\xi^{(l,m,k,p)}$ takes the form
\begin{equation}
\begin{gathered}
\label{6.39}\xi^{(l,m,k,p)}=r_{1/6}^{(l,m,k,p)} z^{1/6}
+\sum_{n=1}^{\infty}r^{(l,m,k,p)}_{(1-7n)/6}\,z^{(1-7n)/6},\\
\,\quad\,l=1,2,3;\,\,\quad\, m=1,2;\,\,\quad\,
k=1,2;\,\quad\,p=1,2,3
\end{gathered}
\end{equation}
The computing the coefficient  $r_{-1}^{(l,m,k,p)}$ gives a result
$r_{-1}^{(l,m,k,p)}=1/6$. The expansion of solution with taking into
account three terms is
\begin{equation}
\begin{gathered}
\label{6.39a}\xi^{(l,m,k,p)}=r_{{1/6}}{z^{1/6}}+\frac16\,{z}^{-1}+
\left( -30\,{g_{{1/6}}}^{4}-9\,{
10}^{2/3}{g_{{1/6}}}^{2}+\right.\\
+150\,{g_{{1/6}}}^{2}c_{{1/3}}-35\,c_{{1/3}}{
10}^{2/3}-30\,{r_{{1/6}}}^{2}{g_{{1/6}}}^{2}+7\,{r_{{1/6}}}^{2}{10}^{2
/3}-\\
-\left.60\,{g_{{1/6}}}^{3}r_{{1/6}}+6\,r_{{1/6}}g_{{1/6}}{10}^{2/3}
 \right)  \left( {10}^{2/3}-2\,{g_{{1/6}}}^{2} \right) ^{-1}  \\
 \left( 6
\,{g_{{1/6}}}^{2}-10\,c_{{1/3}}+3\,{r_{{1/6}}}^{2}+8\,g_{{1/6}}r_{{1/6
}} \right) ^{-1}{g_{{1/6}}}^{-1}{z}^{-{\frac {13}{6}}}
\end{gathered}
\end{equation}
The exponential addition $v^{(l,m,k,p)}(z)$ to solutions
$u^{(l,m,k)}(z)$ is
\begin{equation}
\begin{gathered}
\label{6.40}v^{(l,m,k,p)}(z)=C_2\,z^{1/6} \exp \left[\frac67\,
r^{(l,m,k,p)}_{1/6}\,z^{7/6} + \sum^{\infty}_{n=2} \frac{6}{7(1-n)}
r^{(l,m,k,p)}_{(1-7n)/6}
z^{7(1-n)/6}\right],\\
l=1,2,3;\,\quad\, m=1,2;\,\quad\, k=1,2;\,\quad\,p=1,2,3
\end{gathered}
\end{equation}
Solutions $v^{(l,m,k,p)} (z)$ seem to be divergent ones too.

\section{\!\!\!\!\!\!.\,\, Exponential additions of the third level.}

Let's compute the exponential additions of the third level
$y^{(s)}$, i.e. the additions to the solutions $v^{(l,m,k,p)}(z)$.
The reduced equation for addition $y^{(s)}$ is
\begin{equation}
\label{6.41}M_{s}^{(3)} (z) y^{(s)}=0
\end{equation}
Operator $M_{s}^{(3)}$ is the first variation of \eqref{6.32}.
Equation \eqref{6.41} for $y=y^{(l,m,k,p,s)}$ takes the form
\begin{equation}
\begin{gathered}
\label{6.42}y_{zz} + 3\xi_{z} y+ 3\xi y_{z} +
3\xi^2\,y+4\,\zeta\,y_z+8\,\xi\,\zeta\,y+\\
+6\zeta_z\,y+6\zeta^2\,y-10\,\varphi^{(l)}\,y=0
\end{gathered}
\end{equation}
Using the substitute
\begin{equation}
\label{6.43}\frac{d \ln y}{dz}=\eta
\end{equation}
we obtain
\begin{equation}
\label{6.44}\frac{dy}{dz}=\eta y,\,\,\quad\, \frac{d^2y}{dz^2}
=\eta_z y+\eta^2 y
\end{equation}
From \eqref{6.44} we have equation
\begin{equation}\begin{gathered}
\label{6.45}\eta_{z} +\eta^2+ 3
\xi_z+3\xi\eta+3\xi^2+4\,\eta\,\zeta+8\,\xi\,\zeta+6\zeta_z+6\zeta^2-10\,\varphi^{(l)}=0
\end{gathered}\end{equation}
Monomials of equation \eqref{6.45} is corresponded by points
\begin{equation}
\begin{gathered}
\label{6.46}M_1=(-1,\,1),\,\,\, M_2=(0,\,2),\,\,\, M_3=(-\frac56,\,0),\,\,\, M_4=\left(\frac{1}{6},\,1\right),\\
M_5=\left(\frac13,\,0\right),\,\,\,M_6=\left(\frac16,\,0\right),\,\,\,M_7=\left(\frac13,\,0\right),\,\,\,
M_8=\left(-\frac56,\,0\right),\,\,\,\\
M_9=\left(\frac13,\,0\right),\,\,\,M_{10}=\left(\frac13,\,0\right),\,\,\,\,.\,.\,.
\end{gathered}
\end{equation}
The carrier of equation \eqref{6.45} is formed by points
\eqref{6.46}. The convex set forms the strip, which is similar to
the strip, represented at fig. 3. It should examine edge
$\Gamma_1^{(1)}$, which is passing through points

\begin{equation}
\begin{gathered}
\label{6.47}Q_1=\left(\frac13,0\right),\,\,\, Q
_2=\left(\frac16,1\right),\,\,\, Q_3=\left(0,2\right)
\end{gathered}
\end{equation}
The reduced equation, corresponding to this edge, is
\begin{equation}
\begin{gathered}
\label{6.48}\eta^2 +
3\xi\eta+4\,\eta\,\zeta+8\,\xi\,\zeta+6\,\zeta^{2}-10\,\varphi^{(i)}=0
\end{gathered}
\end{equation}
The basis of the lattice, corresponding to the carrier of equation
\eqref{6.47}, is
\begin{equation*}
\begin{gathered}
\label{6.49}B_1=(1,1),\,\,\,\, B_2=\left(\frac76,0\right)
\end{gathered}
\end{equation*}
The solutions of equation \eqref{6.48} takes the form
\begin{equation}
\begin{gathered}
\label{6.50}\eta^{(l,m,k,p,s)}=q^{(l,m,k,p,s)}\, z^{1/6}\\
 l=1,2,3;\,\,\,\, m,k=1,2; \,\,\,\,p=1,2,3;\,\,\,\,s=1,2;
\end{gathered}
\end{equation}
where $q^{(l,m,k,p,s)}=q$ are the roots of equation
\begin{equation}
\begin{gathered}
\label{6.51}q^2 +3\,q\,
r_{1/6}^{(l,m,k,p)}+4\,q\,g_{1/6}^{(l,m,k)}+8\,r_{1/6}^{(l,m,k,p)}\,g_{1/6}^{(l,m,k)}+\\+
6\,{g_{1/6}^{(l,m,k)}}^2-10\,c_{1/3}^{(l)}=0
\end{gathered}
\end{equation}
The roots of equation \eqref{6.51} are
\begin{equation}
\begin{gathered}
\label{6.51a}
q_{1/6}^{(l,m,k,p,s)}=-\frac32\,r_{1/6}^{(l,m,k,p)}-2\,g_{1/6}^{(l,m,k)}+\\
+(-1)^{s-1}\,\left({\frac94\,{r_{1/6}^{(l,m,k,p)}}^{2}-2\,{r_{1/6}^{(l,m,k,p)}}\,g_{1/6}^{(l,m,k)}-2\,
{g_{1/6}^{(l,m,k)}}^{2}+10\,c_{1/3}^{(l)}}\right)^{(1/2)},\\
l=1,2,3;\,\,\,\,m,k=1,2;\,\,\,\,p=1,2,3;\,\,\,\,s=1,2;
\end{gathered}
\end{equation}

The set of carriers of expansions for  solution $\mathbf{K}$
coincides with \eqref{6.21}. The expansion of solution for
$\eta^{(l,m,k,p,s)}$ takes the form
\begin{equation}
\begin{gathered}
\label{6.52}\eta^{(l,m,k,p,s)}=q^{(l,m,k,p,s)}_{1/6} z^{1/6}
+\sum_{n=1}^{\infty}q^{(l,m,k,p,s)}_{(1-7n)/6}\,z^{(1-7n)/6},\,\,\,\\
l=1,2,3;\,\,\quad\, m=1,2;\,\,\quad\,
k=1,2;\,\,\quad\,p=1,2,3;;\,\,\quad\,s=1,2;
\end{gathered}
\end{equation}
Coefficients $q^{(l,m,k,p,s)}_{1/6},\,\,\,s=1,2$ are determined by
formulas \eqref{6.51a}. The computing of the coefficient
$q^{(l,m,k,p,s)}_{-1}$ gives a result $q^{(l,m,k,p,s)}_{-1}=1/6$.
Exponential addition $y^{(s,p,l,m,k)}(z)$ to the solutions
$v^{(l,m,k,p)}(z)$ is
\begin{equation}
\begin{gathered}
\label{6.53}y^{(l,m,k,p,s)}(z)=C_3\,z^{1/6}\\
\exp \left[\frac67\, q^{(l,m,k,p,s)}_{1/6}\,z^{7/6} +
\sum^{\infty}_{n=2} \frac{6}{7(1-n)} q^{(l,m,k,p,s)}_{(1-7n)/6}
z^{7(1-n)/6}\right]\\
l=1,2,3;\,\quad\, m=1,2;\,\quad\,
k=1,2;\,\quad\,p=1,2,3;\,\quad\,s=1,2
\end{gathered}
\end{equation}

Thus we find three levels of the exponential additions to the
expansions for solutions of equation near point $z=\infty$. Solution
$w(z)$ at $z\rightarrow\infty$ with taking into account the
exponential additions has the expansion
\begin{equation}
\begin{gathered}
\label{6.54}w(z)=c^{(l)}_{1/3}
z^{1/3} -\frac{1}{18z^{2}}+ \sum^{\infty}_{n=2} c^{(l)}_{(1-7n)/3}\, z^{{(1-7n)}/{3}}+\\
+C_1\,z^{-1/4}\,
\exp\{F_1(z)+C_2\,z^{1/6}\exp\{F_2(z)+C_3\,z^{1/6}\exp\{F_3(z)\}\}\}
\end{gathered}
\end{equation}
where $c_{1/3}^{(l)}$ can be computed  by formulas \eqref{1.61},
\eqref{1.62} and \eqref{1.63}; $F_1(z)=F_1^{(l,m,k)}(z)$,
$F_2(z)=F_2^{(l,m,k,p)}(z)$ and $F_3(z)=F_3^{(l,m,k,p,s)}(z)$,
($l=1,2,3;\,\,\,m,k=1,2;\,\,\,p=1,2,3;\,\,\,s=1,2$) are
\begin{equation}
\begin{gathered}
\label{6.55}F_1^{(l,m,k)}(z)=\frac67\, g^{(l,m,k)}_{1/6}\,z^{7/6} +
\sum^{\infty}_{n=2} \frac{6}{7(1-n)} g^{(l,m,k)}_{(1-7n)/6}
z^{7(1-n)/6}
\end{gathered}
\end{equation}
\begin{equation}
\begin{gathered}
\label{6.56}F_2^{(p,l,m,k)}(z)=\frac67\,
r^{(l,m,k,p)}_{1/6}\,z^{7/6} + \sum^{\infty}_{n=2} \frac{6}{7(1-n)}
r^{(l,m,k,p)}_{(1-7n)/6} z^{7(1-n)/6}
\end{gathered}
\end{equation}
\begin{equation}
\begin{gathered}
\label{6.57}F_3^{(l,m,k,p,s)}(z)=\frac67\,
q^{(l,m,k,p,s)}_{1/6}\,z^{7/6} + \sum^{\infty}_{n=2}
\frac{6}{7(1-n)} q^{(l,m,k,p,s)}_{(1-7n)/6} z^{7(1-n)/6}
\end{gathered}
\end{equation}
Coefficients $g^{(l,m,k)}_{1/6}$, $r^{(l,m,k,p)}_{1/6}$ and
$q^{(l,m,k,p,s)}_{1/6}$ are defined by formulas \eqref{6.18},
\eqref{6.19}, \eqref{6.20}, \eqref{6.38a} and \eqref{6.51a}. The
other coefficients are computed sequentially.

\section{\!\!\!\!\!\!.\,\, Summary of the results and  discussion.}

For the solutions of fourth-order analog to the first Painlev\'{e}
equation \eqref{1.7} it is obtained the following expansions.

About a point $z=0$:

1. Four-parameter (with arbitrary constants $c_0,\,\,c_1,\,\,c_2$
and $c_3$) family $G_1^{(0)}1$ of expansion for solution
\eqref{1.27}.

2. Three-parameter (with arbitrary constants $c_1,\,\,c_2$ and
$c_3$) family $G_1^{(0)}2$ of expansion \eqref{1.29}.

3. Two-parameter (with arbitrary constants $c_2$ and $c_3$) family
$G_1^{(0)}3$ of expansion \eqref{1.30}.

4. One-parameter (with arbitrary constant  $c_3$) family
$G_1^{(0)}4$ of expansion \eqref{1.31}.

Families $G_1^{(0)}2$, $G_1^{(0)}3$ and $G_1^{(0)}4$ are the special
cases of family $G_1^{(0)}1$; $G_1^{(0)}3$ and $G_1^{(0)}4$ are the
special cases of $G_1^{(0)}2$; $G_1^{(0)}4$ is the special case of
$G_1^{(0)}3$.

5. Family $G_1^{(1)}1$ of expansion \eqref{1.40} of solution, which
is the special case of families $G_1^{(0)}1$, $G_1^{(0)}2$,
$G_1^{(0)}3$ and $G_1^{(0)}4$.

6. Three-parameter (with arbitrary constants $c_0,\,\, c_3$ and
$c_6$) family $G_2^{(1)}1$ of expansion \eqref{eq1.57d} for solution
of equation \eqref{1.7}.

7. Two-parameter (with arbitrary constants  $c_6$ and $c_8$) family
$G_2^{(1)}2$ of expansion \eqref{eq1.58c} for solution of equation
\eqref{1.7}.

All listed expansions converge for sufficiently small $|z|$.

About a point $z=\infty$:

8. Three expansions $G_3^{(1)}l\,\,(l=1,2,3)$, described by formulas
\eqref{1.61}, \eqref{1.62} and \eqref{1.63}. For each of these
expansions it is found four exponential additions
$G_3^{(1)}lG_1^1mk\,\,(m,k=1,2)$ expressed by formula \eqref{6.23}.
For them it is also computed exponential additions
$G_3^{(1)}lG_1^1mkG_1^{(1)}p\,\,(m,k=1,2;p=1,2,3)$, and then the
proper differential additions  $G_3^{(1)}lG_1^{(1)}mk
G_1^{(1)}pG_1^{(1)}s\,\,(m,k=1,2;p=1,2,3;s=1,2)$ are found too.

The existence and analyticity of expansions, described in items
 1-7, follow from Cauchy
theorem. Families $G_2^{(1)}l\,\,$ and $G_2^{(1)}2$ were first found
in the paper \cite{Kudryashov08}. However the structure of
expansions $G_2^{(1)}1$ and $G_2^{(2)}2$ was not discussed earlier.
The other families of expansions  of solution are found for the
first time.

Comparing the power expansions of equation \eqref{1.7} with power
expansions of Painlev\'{e} equations $P_1 \div P_6$
\cite{Bruno03,Bruno04,Bruno05,Bruno06,Bruno07,Bruno08,Bruno09,Bruno10,Bruno11,Bruno12,Gromak01}
we note, that they differ. This fact can be interpreted as the
additional case for the hypothesis, that the fourth-order equation
\eqref{1.7} determines new transcendental functions just as
equations $P_1 \div P_6$.

\section{\!\!\!\!\!\!.\,\, Appendix. The computation of the basis of the plane lattice.}

Let there is a set $\mathbf{S}$ of points $Q_1,\,\,...,\,\,Q_m$ on
the plane $\mathbb{R}^2$, and there is a zero among them. Our aim is
to compute the basis $B_1,\,\, B_2$ of the minimal lattice
$\mathbf{Z}$, which contains all the points of set $\mathbf{S}$. The
minimality of lattice $\mathbf{Z}$ means that there is no other
lattice $\mathbf{Z} \subset \mathbf{Z}$ and $\mathbf{Z}_1 \neq
\mathbf{Z}$, which also contains set $\mathbf{S}$. The computation
is divided into three steps.

\textit{Step} 1. Let $Q_m=0$, and the others $Q_j \neq 0$. For all
pairs of vectors $Q_j,\,\, Q_k,\,\,1\leq j,\,k<m,j\neq k$ compose
the determinants

\begin{equation}
\label{19_1}\det(Q_jQ_k)\stackrel{def}{=} \Delta_{jk}.
\end{equation}

Among pairs with $\Delta_{jk}\neq 0$ we find one with
$|\Delta_{jk}|= \min |\Delta_{jk}| \neq 0$ in all $j,k=1,\,\,
...,\,\, m-1$. If there are few such pairs, we can take any of them.
Suppose for the sake of simplicity that it is pair $Q_1,\,\,Q_2$.
Other pairs $Q_3,\,\,...,\,\,Q_{m-1}$ are arbitrary ordered.

\textit{Step} 2. Let's find the basis of the lattice, generated by
vectors $Q_1,Q_2,Q_3$. Let $Q_3=a Q_1+ bQ_2$, where $a$ and $b$ are
the rational quantities. Denote integer part of  number $a \in
\mathbb{R}$ as $[a]$ and the fractional part as $\{a\}$, i.e. $\{a\}
= a-[a]$. Denote $Q'_3=\{a\}Q_1 +\{b\}Q_2$. Suppose that
$\min|\det(Q'_3Q_i)|$ for $i=1,2$ reaches  at $i=1$. Then we take
$Q_1$ and $Q'_3$ as the basis vectors and use them to express $Q_2$,
i.e. we get $Q_2=a_1 Q_1 + bQ'_3$. Replace vector $Q_2$ by $Q'_2 =
\{a_1\}Q_1 + \{b_1\}Q'_3$. Among three vectors $Q_1, Q'_2,Q'_3$ we
find the pair with the least modulus of determinant. Using this pair
we distribute the third vector, take his fractional part and so on.
At some step $l$ we obtain that the fractional part of the third
vector equals zero. The latest pair of vectors
$Q^{(l)}_2,\,\,Q^{(l)}_3$ gives the basis of minimal lattice,
containing the points $Q_1,\,\, Q_2,\,\, Q_3$.

\textit{Step} 3. For vectors $Q^{(l)}_2,\,\,Q^{(l)}_3,\,\,Q_4$ we
realize step 2 and get vectors $\Tilde{Q}_3,\,\Tilde{Q}_4$ and so
on. After looking through all $Q_j,\,\,j \leq m-1$, we get the pair
of vectors $Q^*_{m-2},\,\,Q^*_{m-1}$, which is the basis of minimal
lattice, containing the set $\mathbf{S}$.

\textbf{Remark.}  The analogous algorithm allows as to find the
basis of minimal lattice in $\mathbb{R}^n$, containing the given
finite set  $\mathbf{S}$. If $n=1$ it's the Euclid algorithm.

\textbf{Example 8.1.} Let's consider equation \eqref{1.5}. It's
carrier consists of six points \eqref{1.7}. Move them by vector
$Q_4=(1,0)$. We obtain

\begin{equation*}
\label{19_2}Q'_1=(-5,\,\,1),\,\,\, Q'_2=(-3,\,\,2),\,\,\,
Q'_3=(-1,\,\,3),\,\,\, Q'_4=0
\end{equation*}

For vectors $Q'_1,\,\,Q'_2,\,\,Q'_3$ we compute the pairwise
determinants

\begin{equation}
\label{19_3} \Delta_{1 2}=\left| \begin{gathered}-5 \,\,\, 1 \\ -3
\,\,\, 2
\end{gathered} \right|=-7,\,\,\,\, \Delta_{13}\left| \begin{gathered}-5 \,\,\, 1 \\ -1
\,\,\, 3
\end{gathered} \right|=-14,\,\,\,\,\Delta_{23}\left| \begin{gathered}-3 \,\,\, 2 \\ -1 \,\,\, 3 \end{gathered}
\right|=-7.
\end{equation}

Thus as the initial pair we can use  vectors  $Q'_1,\,\,Q'_2$ or
$Q'_2,\,\,Q'_3$. Let's take $Q'_1,\,\,Q'_2$ to fix the idea. We look
for the expansion $Q'_3= aQ'_1 + bQ'_2 =a(-5,1) +b(-3,2)$, for that
we are to solve the linear system of equations

\begin{equation}
\begin{gathered}
\label{19_4} -5a-3b=-1,\\
a+2b=3.
\end{gathered}
\end{equation}

We get $a=-1,\,\,b=2$. As long as $\{a\} =\{b\}=0$, then the vectors
$B_1=Q'_1$ and $B_2=Q'_2$ generate the basis of the lattice of
shifted carrier of equation  \eqref{1.5}.


\begin{thebibliography}{99}


\bibitem{Kudryashov01} \textit{Kudryashov N.A.} The first and second Painlev\'e equations of higher order and
some relations between them // Phys. Letters. A. 1997. V. 224. N 6.
P. 353--360.

\bibitem{Kudryashov02} \textit{Kudryashov N.A.} Analitical theory of nonlinear differential equations, Moscow -- Igevsk,
Institute of Computer research, 2004. 360 p.

\bibitem{Golubev} \textit{Golubev V.V.} Lectures on integration of the equation
of motion of a rigid body about a fixed point, Moscow, Gostekhizdat,
1953. 288 p. (in Russian).

\bibitem{Olver} \textit{Olver P.J.} Hamilton and non-Hamilton models for water waves // Lecture Notes
in Physics. N.Y.: Springer, 1984. N 195. P. 273-290.

\bibitem{Kudryashov03} \textit{Kudryashov N.A.} First integrals of nonlinear wave dynamics // Applied Mathematics and Mechanics, 2005.
v. 69. No. 2. pp. 884-894 (in Russian).

\bibitem{Henon} \textit{H\'enon M., Heiles C.} The applicability of the third integral of motion:
some numerical experiments // Astron. J. 1964. V. 69. N 1. P.
73--79.

\bibitem{Fordy} \textit{Fordy A.P.} The H\'enon --- Heiles system revisited // Physica D. 1991. V. 52. N 2--3. P. 204--210.

\bibitem{Hone} \textit{Hone Andrew N.W.} Non -- autonomous H\'enon --- Heiles system // Physica D. 1998. V. 118 P. 1 -- 16.

\bibitem{Kudryashov04} \textit{Kudryashov N.A.} On new transcendents defined by nonlinear ordinary
differential equations // J. Phys. A.: Math. Gen. 1998. V. 31. N 6.
P. L.129--L.137.

\bibitem{Kudryashov05} \textit{Kudryashov N.A.} Transcendents defined by nonlinear
fourth-order ordinary differential equations // J. Phys. A.: Math.
Gen. 1999. V. 32. N 6. P. 999--1013.

\bibitem{Kudryashov06} \textit{Kudryashov N.A.} Fourth-order analogies to the Painlev\'e equations // J. Phys.
A.: Math. Gen. 2002. V. 35. N 21. P. 4617--4632.

\bibitem{Kudryashov07} \textit{Kudryashov N.A.} Amalgamations of the Painlev\'e equations // J. Math.
Phys., 2003, V. 44. N 12. P. 6160--6178

\bibitem{Kudryashov08} \textit{Kudryashov N.A., Soukharev M.B.} Uniformization and Transendence of solutions for
the first and second Painleve hierarchies, Physics Letters A v. 237,
1998, 206-216

\bibitem{Kudryashov09} \textit{Kudryashov N.A., Soukharev M.B.} Discrete equations corresponding to fourth - order
differential equations of the P2 and K2 hierarchies, ANZIAM,
Industrial and Applied Mathematics, 2002

\bibitem{Kudryashov10} \textit{Kudryashov N.A.} Nonlinear differentil equations of the fourth order with solutions in
the form of transcendents, Theoretical and mathematical physics, v.
122, No. 1, 2000, 72 - 86 (in Russian)

\bibitem{Kudryashov11} \textit{Kudryashov N.A., Pickering A.} Rational solutions
for Schwarzian integrable hierarchies, Journal of Physics A, Math.
Gen., v. 31, 1998, 999 - 1014

\bibitem{Joshi01} \textit{Creswell G., Joshi N.} The discrete first,second and thirty - fourth Painleve hierarchies,
Journal of Physics A, Math. Gen., v. 32, 1999, 655 - 669

\bibitem{Mugan01} \textit{Mugan U., Jrad F.} Painleve test and the first Painleve hierarchy,
Journal of Physics A, Math. Gen., v. 32, 1999, 7933 - 7952

\bibitem{Cosgrove01} \textit{Cosgrove C.M.} Higher - order Painleve equations in the polynomial class. I. Bureau
symbol P2, Study Appl. Math., v. 104, 2000, 1- 65

\bibitem{Gordoa01} \textit{Gordoa P.R.} Backlund transformations for the second member of the first Painleve hierarchy,
Physics Letters A, v. 287, 2001, 365 - 370

\bibitem{Pickering01} \textit{Pickering A.} Coalescence limits for higher order Painleve equations,
Physics Letters A, v. 301, 2002, 275 - 280

\bibitem{Clarkson01} \textit{Clarkson P.A., Hone A.N.W., Joschi N.} Hierarchies of difference equations
and Backlund transformations, Journal of Nonlinear Mathematical
physics, v. 10, 2003

\bibitem{Kawai01} \textit{Kawai T., Koike T., Nishikawa Y., Takei Y.} On the complete description of the
Stokes geometry for the first painleve hierarchy, 2004,
preprint/RS/RIMS 1471, Kioto

\bibitem{Bruno01} \textit{Bruno A.D.} Power geometry in algebraic and differential equations,
Moscow, Nauka, Fizmatlit, 1998, 288 p. (in Russian).

\bibitem{Bruno02} \textit{Bruno A.D.} Asimptotics and expansions of solutions for ordinary differential
equation, Uspekhi of Mathematical sciences, v.59, No. 3, 2004, pp.
31-80 (in Russian).

\bibitem{Bruno03} \textit{Bruno A.D., Petrovich V.Yu.} Particularities of solutions of the first Paileve
equation, Moscov,  Preprint Keldysh Institute of Applied
Mathematics, No.75, 2004,(in Russian).

\bibitem{Bruno04} \textit{Bruno A.D.} Power Geometry as a new calculus, in book H.G.W. Begehr
et al (eds) Analysis and Applications, ISAAC, 2001, 51-71, 2003
Kluwer

\bibitem{Bruno05} \textit{Bruno A.D., Chukhareva I.B.} Power expansions of the sixth Painleve
equation, Moscow, Preprint Keldysh Institute of Applied Mathematics,
No.49, 2003 (in Russian).

\bibitem{Bruno06} \textit{Bruno A.D., Karulina E.S.} Power expansions of the fifth Painleve
equation, Moscow, Preprint Keldysh Institute of Applied Mathematics,
No. 50, 2003.

\bibitem{Bruno07} \textit{Bruno A.D., Zavgorodnaya Yu.V.} Power series and non -- power asimptotics
of the second Painleve equation, Moscow, Preprint Keldysh Institute
of Applied Mathematics, No. 48, 2003.

\bibitem{Bruno08} \textit{Bruno A.D., Gridnev A.V.} Power and exponential expansions of the third
Painleve equation, Moscow, Preprint Keldysh Institute of Applied
Mathematics,  No. 51, 2003.

\bibitem{Bruno09} \textit{Bruno A.D., Karulina E.S.} Power expansions of the fifth Painleve
equation, Reports of Russian Academy Sciences, 2004, v. 395, No. 4,
pp. 439-444.

\bibitem{Bruno10} \textit{Bruno A.D., Goruchkina I.B.} Power expansions of the sixth Painleve
equation,Reports of Russian Academy Sciences, v. 395, 2004, No. 6,
pp. 733-737.

\bibitem{Gromak01} \textit{Gromak V.I., Laine I., Shimomura S.} Painleve Differential Equations in
the Complex Plane, Walter de Gruyter, Berlin, New York, 2002.



\end{thebibliography}
\end{document}